\documentclass[%
 reprint,
 amsmath,amssymb,
 aps,
]{revtex4-2}

\usepackage{graphicx}
\usepackage{dcolumn}
\usepackage{bm}
\usepackage{booktabs}
\usepackage{array}
\usepackage{color,soul}
\usepackage{chngcntr}
\usepackage{subfigure}

\begin{document}


\title{Evaluating ion dynamics through Coulomb and Yukawa interaction potentials in one component strongly coupled plasmas}

\author{Swati Swagatika Mishra}
\email{swatis@iitk.ac.in}
\author{Sudeep Bhattacharjee}%
 \email{sudeepb@iitk.ac.in}
\affiliation{%
 Department of Physics, Indian Institute of Technology Kanpur, Uttar Pradesh 208016, India
}%


\author{Pascal Brault}
 \email{pascal.brault@univ-orleans.fr}
\affiliation{
GREMI UMR 7344 CNRS Universite dOrleans, Cedex 2 France}%

\date{\today}

\begin{abstract}
Atmospheric pressure helium plasmas are investigated through molecular dynamics simulations at room temperature (300 K) for various ionization fractions ($\chi_i = 10^{-1} - 10^{-5}$) in the strongly coupled regime (ion coupling parameter, $\Gamma_i \sim 1 - 10$) employing Coulomb and Yukawa interaction potentials. The role of electron screening in ion dynamics and energetics is examined through ion and gas temperatures, mean squared displacement of ions, ion coupling parameter, and radial distribution function of the system. It is found that electron screening in the Yukawa potential significantly limits the disorder-induced heating (DIH) mechanism for strongly ionized plasmas ($\chi_i$ $\ge 10^{-3}$). Whereas, ions show a prominent sub-diffusive behavior associated with the DIH during the non-equilibrium phase for the Coulomb potential. The DIH mechanism is explained using a model based upon the conservation of energy. However, for weakly ionized plasmas ($\chi_i \le 10^{-4}$), the maximum ion temperatures are almost similar for both potentials. Furthermore, electron screening affects the separation distance and arrangement of the ion-neutral pairs for all the values of $\chi_i$. In general, Yukawa potential results in a lower mean potential energy of the interacting particles, which is energetically favourable for the stability of the system.
 
\end{abstract}



\maketitle
\section{Introduction}
Plasmas at atmospheric pressure have been studied extensively in a wide range of experiments and found their applications in surface processing and biomedical fields \cite{barman2023improving,behmani2023plasma,behmani2024frequency,neyts2017molecular}. The criticality of their research lies in the fact that it has led to the emergence of a new field known as plasma medicine \cite{bernhardt2019plasma,keidar2018plasmas}. These plasmas are often generated in laboratories at room temperatures employing RF sources \cite{hofmann2011power}, microwaves \cite{ouyang2012characterization}, DC sources \cite{wang2006simulation}, nano-second pulsed discharges \cite{van2012time}, and dielectric barrier discharges \cite{barman2020characteristics}. Depending on the mode of generation, the value of ionization fraction ($\chi_{i} = n_i/(n_i + n_n) $, where $n_i$ and $n_n$ are the ion and gas number densities, respectively) in the atmospheric pressure plasmas, ranges from $10^{-1}$ to $10^{-8}$, and divides the plasma into strong, moderate and weakly ionized regimes. Based on their ionization fraction, the plasmas show significant heating and strong coupling among the ions after their generation \cite{acciarri2022strong, wang2006simulation, kothnur2003structure}, where the ion coupling parameter ($\Gamma_i$), which is the ratio of mean ion potential energy to the mean ion kinetic energy becomes $\geq 1$. 
In strongly ionized plasmas, the ion heating is assumed to be caused by the disorder-induced heating (DIH) mechanism \cite{acciarri2022strong}. DIH arises as the ions tend to move to a lower potential energy state from a disordered state and release kinetic energy in the process. However, for weakly ionized plasmas, the heating is often attributed to the discharge geometry and type of plasma source \cite{wang2006simulation,kothnur2003structure}. 
For the moderate and weakly ionized plasmas, most of the simulation and theoretical efforts are focused on understanding the plasma chemistry, electron temperature ($T_e$), and electron densities ($n_e$) for a given plasma geometry and generation scheme \cite{wang2006simulation,kothnur2003structure,kushner2005modelling}. Furthermore, the contributions of intra and inter-species interactions and internal structures of the bulk plasmas in plasma heating and ion coupling strength are not very well understood, even for the strongly ionized plasmas. In this context, we employ molecular dynamics (MD) simulations to investigate the ion and neutral gas dynamics and their implications in ion energies, ion coupling strength, ion and neutral temperatures, and the internal structures in an atmospheric pressure helium plasma for $\chi_{i}$ (= $10^{-1}$ to $10^{-5}$), at room temperature (300 K), thus covering the strong, moderate and weakly ionized plasma regimes. MD simulations efficiently track each particle's trajectories and easily convey molecular-level information \cite{thompson2022lammps,mishra2022temperature,mishra2024collisional,ramkumar2024uncertainties}. However, while employing the MD simulations, it is imperative to designate the intra and inter-species interactions with appropriate potentials. Therefore, in this work, we discuss the performance of both Coulomb and Yukawa potentials as the probable ion-ion interaction potential for atmospheric pressure plasmas, which may help us to decide the appropriate potential, depending upon $\chi_i$ and other plasma parameters. Section II of this article discusses the inter-particle potentials in-depth, along with the simulation details, followed by the important results outlined in section III and a discussion in section IV. The article is concluded in section V.

\vspace{2cm}
  
\section{Simulation details}

Typically, the plasmas consist of neutrals, ions, and electrons, and in atmospheric pressure plasmas, the gas temperature ($T_g$) $\sim$ ion temperature ($T_i$), and the electron temperature, $T_e$ $\gg$ $T_i$. 
Therefore, in most plasma simulations, one component plasma (OCP) model is used \cite{arkhipov2017direct,dharuman2018controllable, acciarri2022strong}, and electrons are not explicitly considered in the simulation system.
Commonly, for atmospheric pressure plasmas, ionic interactions (which are singly charged He ions in the study) are depicted by the Coulomb potential \cite{acciarri2022strong,acciarri2023influence}, given by 
\begin{equation}
    \phi_{C} (r) = \frac{1}{4\pi\epsilon_{0}}\frac{q^2}{r},
\end{equation}
where $\epsilon_0$ is the permittivity of free space, $q$ is the charge of the ions, and $r$ is the separation distance between a pair of ions. This model is seen to work well for quantifying the DIH for strongly ionized argon plasmas with $\chi_i \ge 10^{-2}$ \cite{acciarri2022strong}. In the present simulation work, this model will be tested for He plasmas at various ionization fractions. However, in Coulomb potential, the effect of the screening provided by the background electrons is fully neglected, which may have a role in the ion dynamics, as seen in the case of ultracold neutral plasmas \cite{lyon2013limit}. Therefore, the screened Coulomb (Yukawa) potential, given by
\begin{equation}
    \phi_{Y} (r) = \frac{1}{4\pi\epsilon_{0}}\frac{q^2 e^{-\kappa r}}{r},
\end{equation}
is also employed to quantify the screening effect at different values of $\chi_{i}$. Here, $\kappa$ is the screening constant due to the electrons and is calculated assuming the ions and the electrons have the same densities ($n_i = n_e$), using the relation, 
\begin{equation}
\kappa^2 = \frac{ n_e q^2}{\epsilon_{0} k_B T_e}. 
\end{equation}
Furthermore, the electron temperature ($T_e$) is chosen to be 1 eV (11600 K) in this work, which is a typical value of $T_e$ in atmospheric pressure plasmas \cite{wang2006simulation}.  
The values of $\kappa$ are tabulated in Table 1 along with the values of mean inter-ionic distances ($a_{ii}$) and the ion plasma frequencies ($\omega_{pi} = \sqrt{e^2n_i/\epsilon_0 m_i}$, where $e$ is the elementary charge and $m_i$ is the mass of the ion) for different $\chi_i$. It is observed that for lower values of $\chi_{i}$, the values of $\kappa \sim 10^{-4}$ \AA$^{-1}$. Therefore, it is expected that the effect of electron screening diminishes for weakly ionized plasmas. The inverse of $\kappa$, known as the Debye radius ($r_d$), is $> a_{ii}$ for all the values of $\chi_i$, to ensure effective screening within the Debye sphere \cite{bittencourt2013fundamentals,chen1984introduction}. The Coulomb potential calculations are carried out using the particle-particle particle-mesh (P3M) solver to incorporate the long-range effects. However, for Yukawa potential, a cut-off length is chosen depending on the value of $a_{ii}$ such that the cut-off length $>$ $a_{ii}$ and the magnitude of the force at the cut-off length becomes $10^{-10}$ times of the maximum force value. 

\begin{table}
\centering
\caption{\label{tab:table3} {Simulation parameters for different ionization fractions ($\chi_i$).} }
\begin{tabular}{cccc}
\toprule
$\chi_i$ & $\kappa (\times 10^{-3}$ \AA$^{-1}$)& $a_{ii}$ (\AA) & $\omega_{pi}$ ($\times 10^{12}$rad/s$^{-1}$) \\ \hline
 \midrule
 $10^{-1}$&21.03& 46.04&1.0336\\
$10^{-2}$&6.65& 99.19&0.3268 \\
 $10^{-3}$&2.1& 213.72&0.1034 \\
 $10^{-4}$&0.665& 460.43&0.0326\\
$2\times10^{-5}$&0.21& 787.32&0.0051\\
 \hline
\end{tabular}
\end{table}

The plasma properties are also expected to be influenced by the neutral gas. Therefore, ion-neutral and neutral-neutral interactions are incorporated in the simulation. The ion-neutral interactions are guided by the modified charge-induced dipole potential \cite{acciarri2022strong}, given by
\begin{equation}
    \phi_{ind} (r) = \frac{q^2}{8\pi\epsilon_{0}}\frac{\alpha_R a_B^3}{r^4}(\frac{r_{\phi}^8}{r^8} - 1),
\end{equation}
where, $\alpha_R$ is the relative polarizability for He (= 1.38 \cite{puchalski2020qed}), $a_B$ is the Bohr's radius, and $r_\phi$ is the radius at which the repulsive core acts. 
The parameter $r_\phi$ (= $c a_{in}$, where $a_{in}$ is the interparticle separation distance for an ion and neutral pair, and $c$ is a multiplicative constant) is calibrated using similar techniques followed by Acciarri et al. \cite{acciarri2022strong}.
For $T_g$ = 300 K, $c$ is found to be 0.133. The cut-off length is set to $\sim $ 10 $a_{in}$. 

The neutral-neutral interactions are modeled by Lennard-Jones potential given by, 
\begin{equation}
    \phi_{LJ} (r) = 4\epsilon [(\frac{\sigma}{r})^{12} - (\frac{\sigma}{r})^6],
\end{equation}
where $\epsilon $ is the energy parameter (0.00094 meV for He), and $\sigma$ is the length parameter (2.64 {\AA} for He) \cite{mishra2022temperature}. The cut-off length is chosen to be 50 \AA.

All the MD simulations are carried out using the LAMMPS software package \cite{thompson2022lammps}. 
The maximum number of ions ($N_i$) is 10,000 for $\chi_i = 10^{-1}$, whereas for $\chi_{i} \leq 10^{-4}$, $N_i$ = 100 is fixed. The system sizes ($L$) are chosen in such a way that atmospheric gas pressure is maintained and $L \gg r_d$, which ensures an important plasma criterion \cite{bittencourt2013fundamentals,chen1984introduction}.
The neutral He particles are distributed randomly in the system and assigned with an initial Gaussian velocity distribution corresponding to the mean temperature $T_g$ = 300 K. The He atoms are then equilibrated at constant temperature in a canonical (NVT, where particle number ($N$), volume ($V$), and temperature ($T$) are constant) ensemble using a Nose-Hoover thermostat \cite{evans1985nose}. After the system is thermally equilibrated, a fraction of He particles are ionized, and the ions are introduced randomly in the system depending on the value of $\chi_i$. The initial velocity distribution of the He ions is of the Gaussian type with the mean temperature corresponding to 300 K. The system is then allowed to evolve in a microcanonical (NVE, where $N$, $V$, and energy ($E$) are constant) ensemble, and $T_i$, $T_g$, and ion energies are recorded. 
As the system evolves, $T_i$ approaches $T_g$, and the thermal equilibrium is achieved when $T_i \sim T_g$ = $T_s$, which is defined as the system temperature.
\begin{figure*}
\begin{subfigure}
  \centering
  \includegraphics[width= 8.15 cm, height = 6.9 cm]{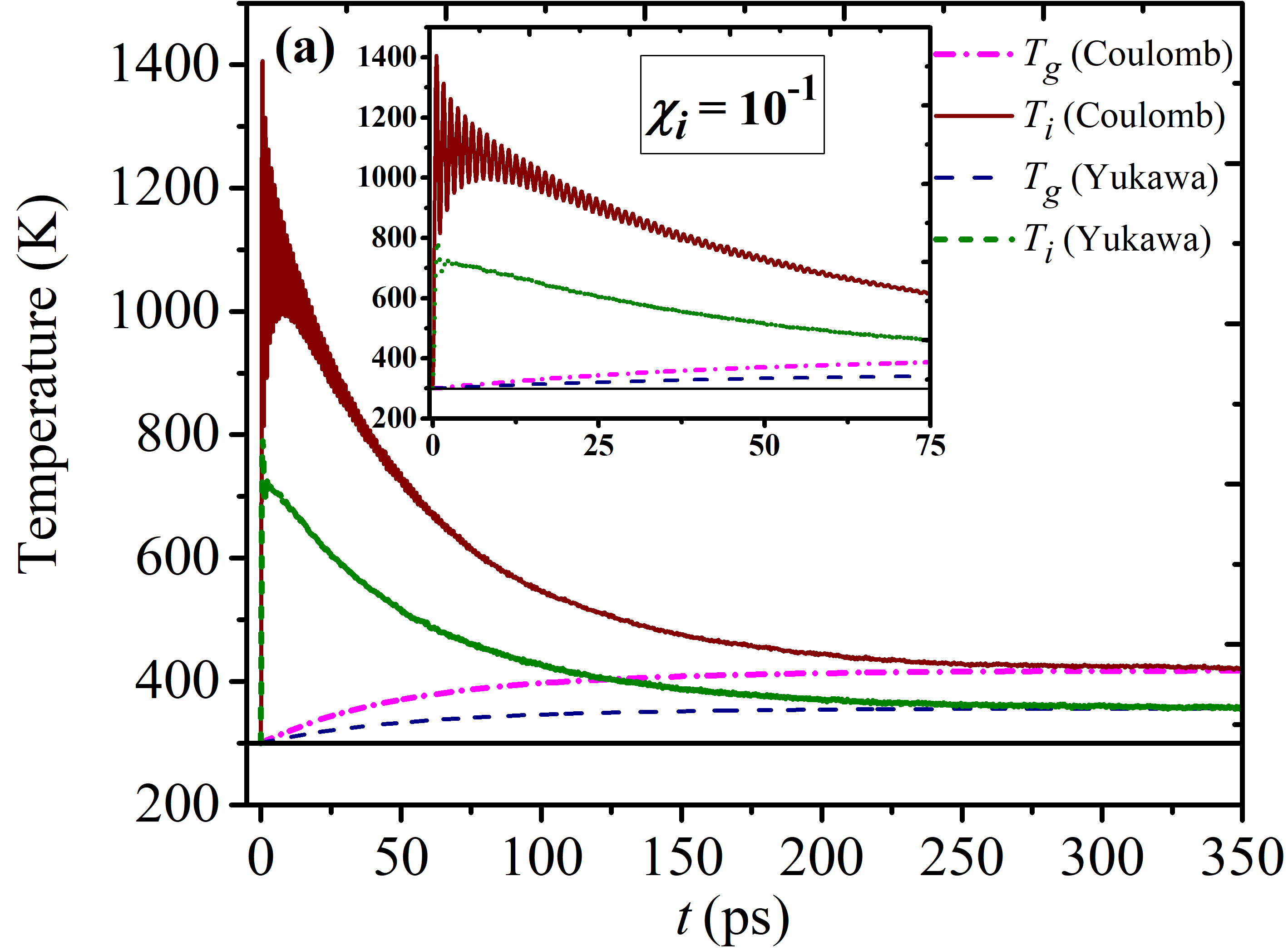}  
  \label{fig:sub-first}
\end{subfigure}
\begin{subfigure}
  \centering
  \includegraphics[width= 8.15 cm, height = 6.9 cm]{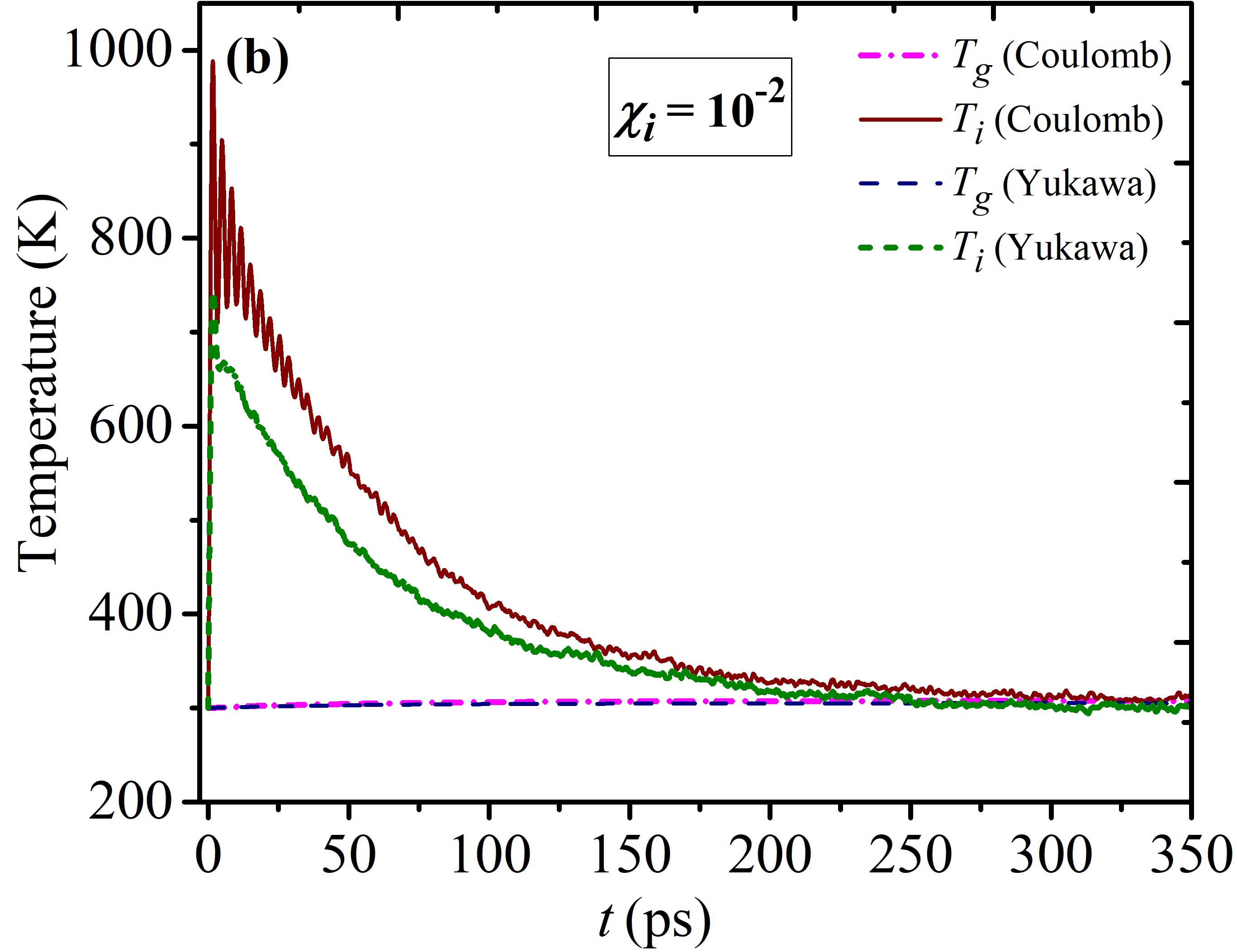}  
  \label{fig:sub-second}
\end{subfigure}
\begin{subfigure}
  \centering
  \includegraphics[width= 8.15 cm, height = 6.9 cm]{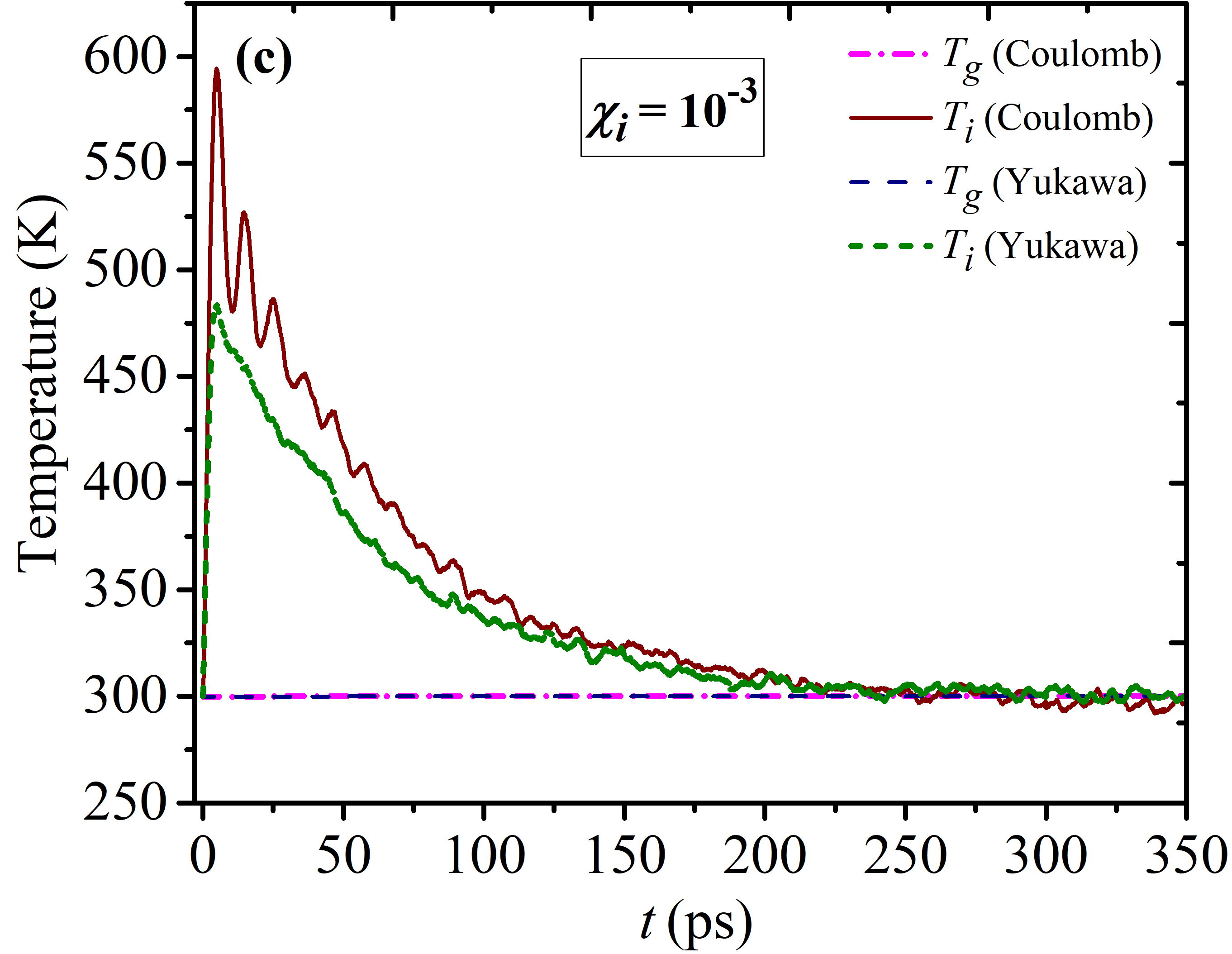}  
  \label{fig:sub-third}
\end{subfigure}
\begin{subfigure}
  \centering
  \includegraphics[width= 8.15 cm, height = 6.9 cm]{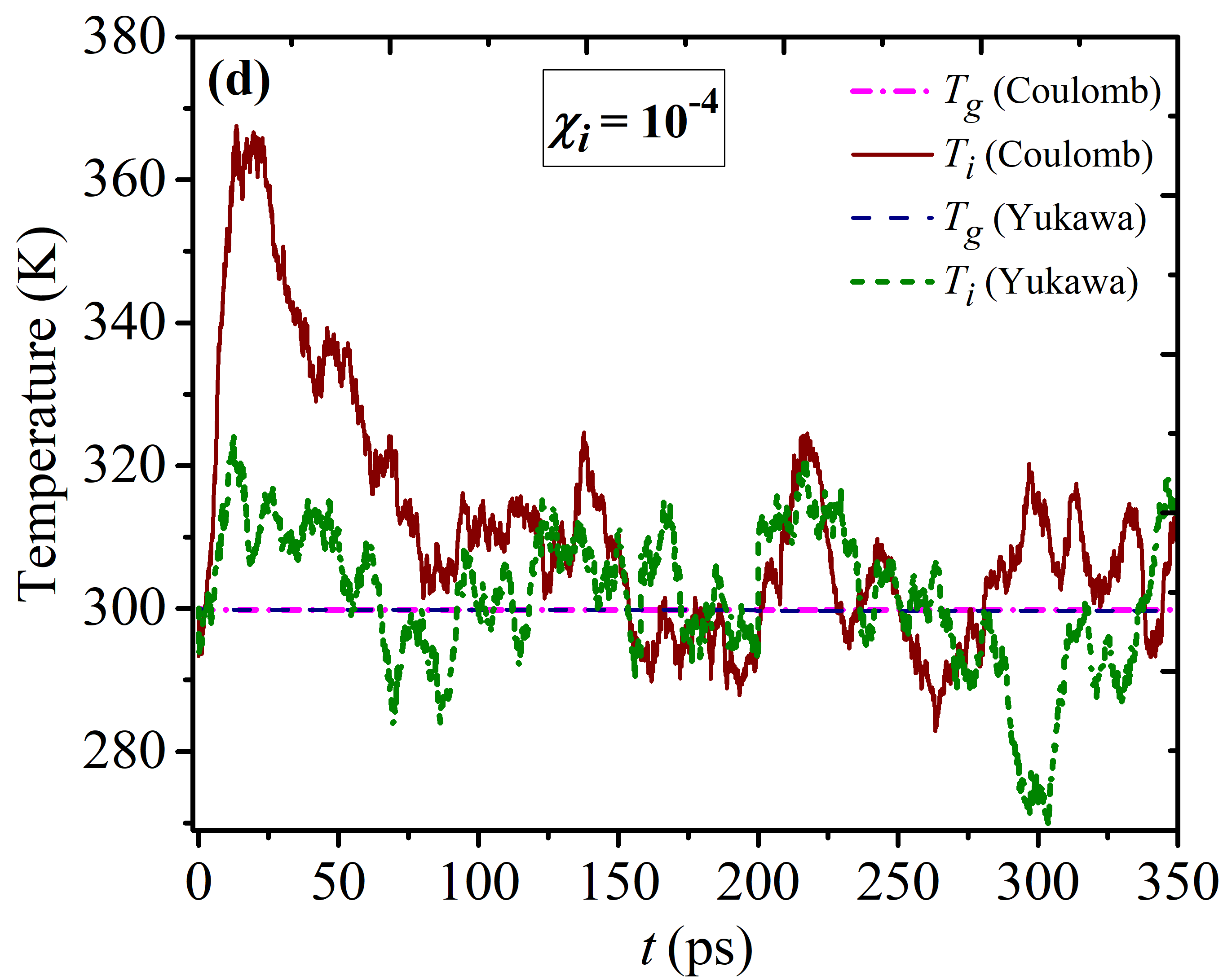}  
  \label{fig:sub-fourth}
\end{subfigure}
\caption{Temporal evolution of ion and gas temperatures ($T_i$ and $T_g$) for different ionization fractions. (a) $\chi_i$ = $10^{-1}$, (b) $\chi_i$ = $10^{-2}$, (c) $\chi_i$ = $10^{-3}$ and (d) $\chi_i$ = $10^{-4}$. The black solid line in Figure 1(a) shows the initial system temperature = 300 K.  }
\label{fig:fig}
\end{figure*}

The simulations are run for five different initial spatial distributions of neutrals and ions for each value of $\chi_i$. The values of $T_i$ and $T_g$ are then averaged and presented in the temperature plots (cf. Figure.1). 

\vspace{1cm}
 
\section{Results}
\subsection{Temperature profiles and ion dynamics }

Figure 1 shows the spatially averaged ion and neutral temperature profiles during the non-equilibrium condition in the system, i.e., $T_i \ne T_g$, for both potentials (Coulomb and Yukawa). The temperatures are sampled during the NVE run of the simulation. The curves are shown till $T_i$ approaches $T_g$ ($\sim 350$ ps). The maximum error in the $T_i$ and $T_g$ curves are 8.12\% and 0.2\%, respectively, which are obtained from the standard deviation errors of the spatially averaged curves. It is observed that the DIH mechanism is prominent for $\chi_{i} \ge 10^ {-3}$, which results in ion heating. $T_g$ shows an increasing trend for $\chi_{i} \ge 10^ {-2}$, which later equilibrates with $T_i$ (cf. Figure 1(a) and (b)). Furthermore, for $\chi_{i} \ge 10^ {-3}$, the $T_i$ curve shows significant oscillations for Coulomb potential. The frequency of the oscillations decreases, whereas the width of each peak increases with a decrease in $\chi_{i}$. The time period within which the oscillations last is quantified in terms of plasma frequency ($f^{-1} = 2\pi/\omega_{pi}$). For $\chi_{i}$ = $10^{-1}$, $10^{-2}$ and $10^{-3}$ the oscillations last for $\sim$ 13$f^{-1}$, 2.6$f^{-1}$, and 0.9$f^{-1}$, respectively. The oscillations last for a longer period for the higher value of $\chi_i$. It shows that the ions interacting with Coulomb potential suffer rapid displacements from their respective mean positions in the potential well formed by the nearest neighbors \cite{castro2009role}, resulting in fluctuating repulsive potential energy that gets converted to the oscillating kinetic energy. 
Furthermore, with a higher number of ions, this phenomenon lasts for longer periods due to higher repulsive energies. In contrast, such oscillations are very much damped for Yukawa potential. For example, for $\chi_i = 10^{-1}$, there is only one secondary peak, which vanishes for lower values of $\chi_i$ ($\le 10^{-3}$). This signifies that the screening due to the electrons weakens the repulsive force between the ions. For weakly ionized plasmas, the peaks gradually vanish, with $\chi_i = 2\times 10^{-5}$ not showing any peaks for any potential (not shown). Rather, the $T_i$ curve oscillates with significant thermal and statistical fluctuations around a mean temperature. It is also observed that for $\chi_{i} \ge 10^{-3}$, the ions attain the maximum temperature ($T_{i_\_max}$) at similar times for both potentials. For example, for $\chi_i = 10^{-1}$, the ions attain $T_{i_\_max}$ = 1405.84 K at $t$ = 0.57 ps for Coulomb potential and $T_{i_\_max}$ = 792.4 K at $t$ = 0.67 ps for Yukawa potential. However, for $\chi_i \le 10^{-4}$, $T_{i_\_max}$ is attained faster for Yukawa potential compared to Coulomb potential. Additionally, with decreasing $\chi_i$, the time corresponding to $T_{i_\_max}$ increases.

Figure 2 shows that the value of $T_{i_\_max}$ decreases with the decrease in $\chi_i$, demonstrating that the DIH mechanism is prominent for $\chi_{i} \ge 10^{-3}$. $T_{i_\_max}$ for a constant $\chi_i$ is higher for Coulomb potential, as seen in Figure 2 for all the values of $\chi_i$, and the values start approaching for both potentials for $\chi_i \le 10^{-4}$. This signifies that ion heating is directly influenced by electron screening.
 The gas temperature at the start of the thermal equilibrium ($T_{g_\_eq}$) for both potentials is also plotted in Figure 2. The values are similar for both potentials for $\chi_i <$ $10^{-2}$.

\begin{figure}
         \centering
         \includegraphics[width= 8.3 cm, height = 6.5 cm]{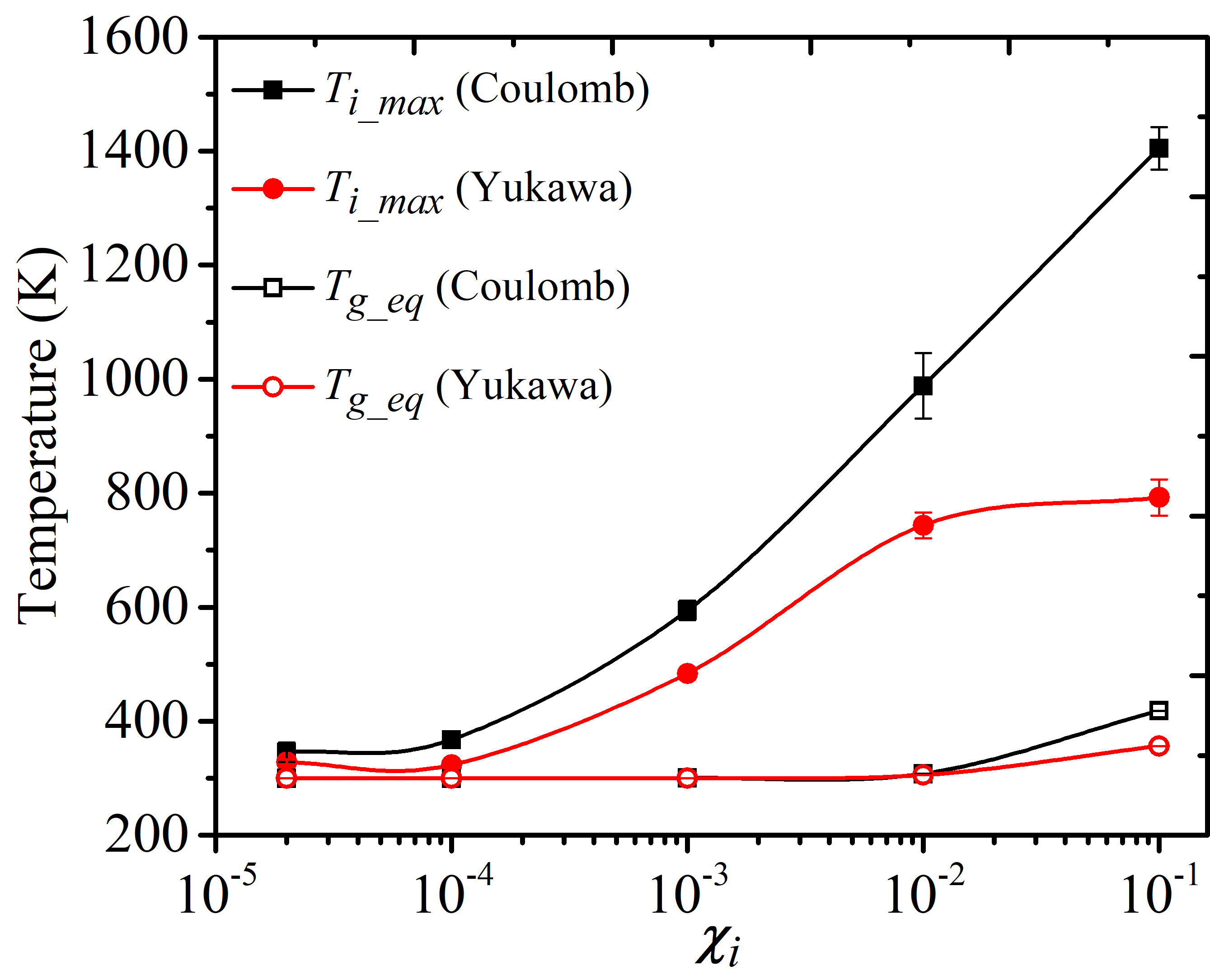}
         \caption{Maximum ion temperature ($T_{i_\_max}$) and equilibrium gas temperature ($T_{g_\_eq}$) versus the ionization fraction ($\chi_{i}$).}
\end{figure}

\begin{figure}
         \centering
         \includegraphics[width= 8.3 cm, height = 6.5 cm]{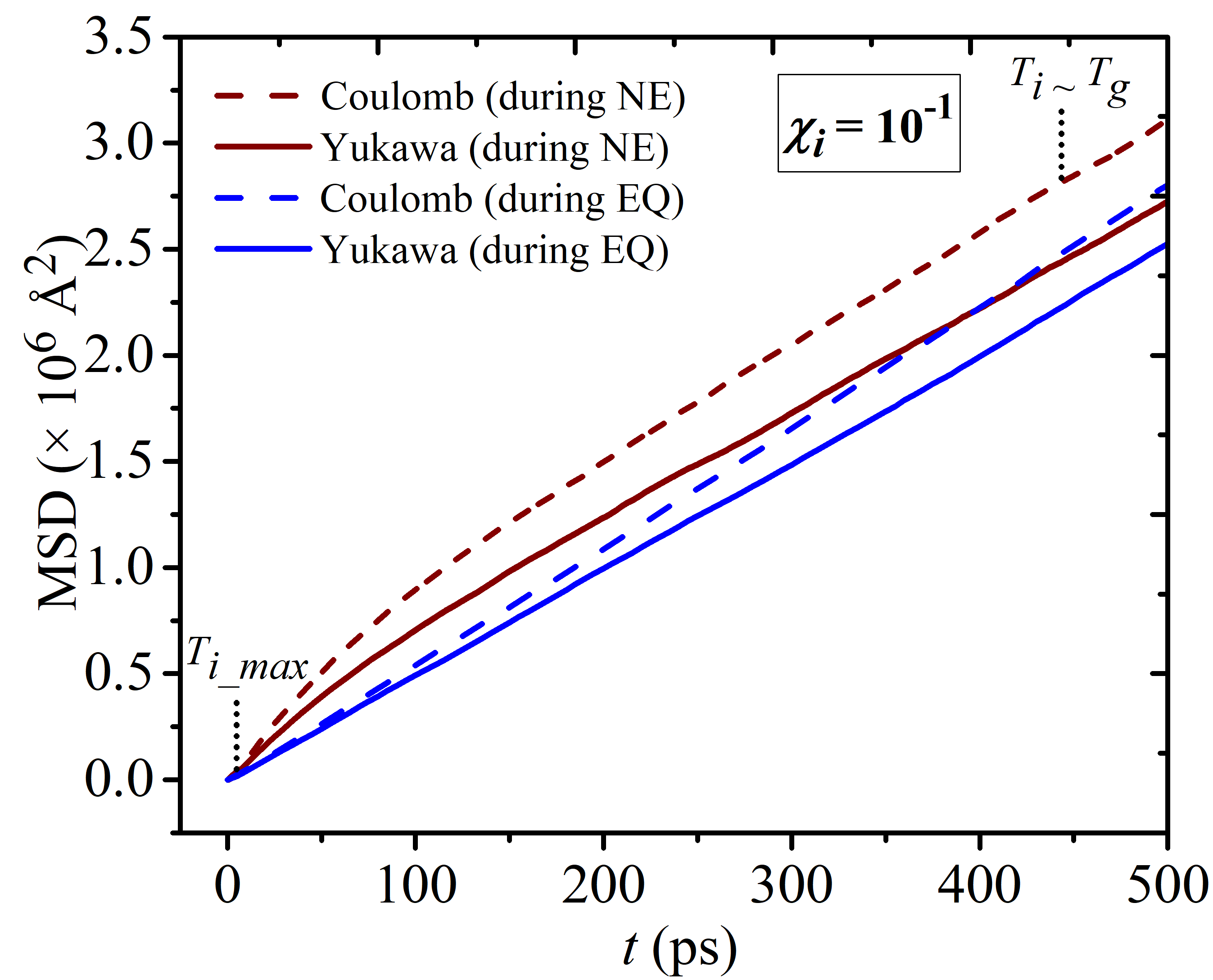}
         \caption{Mean squared displacement for ions (MSD\textsubscript{i}) for $\chi_i = 10^{-1}$. The MSD\textsubscript{i} is calculated during the non-equilibrium (NE) and equilibrium (EQ) phases, with two different origin times. The curve for Coulomb potential (NE) shows markings for different events, such as (i) $T_{i_\_max}$ and (ii) the start of the equilibrium ($T_i \sim T_g$).} 
\end{figure}

\begin{figure}
         \centering
         \includegraphics[width= 8.3 cm, height = 6.5 cm]{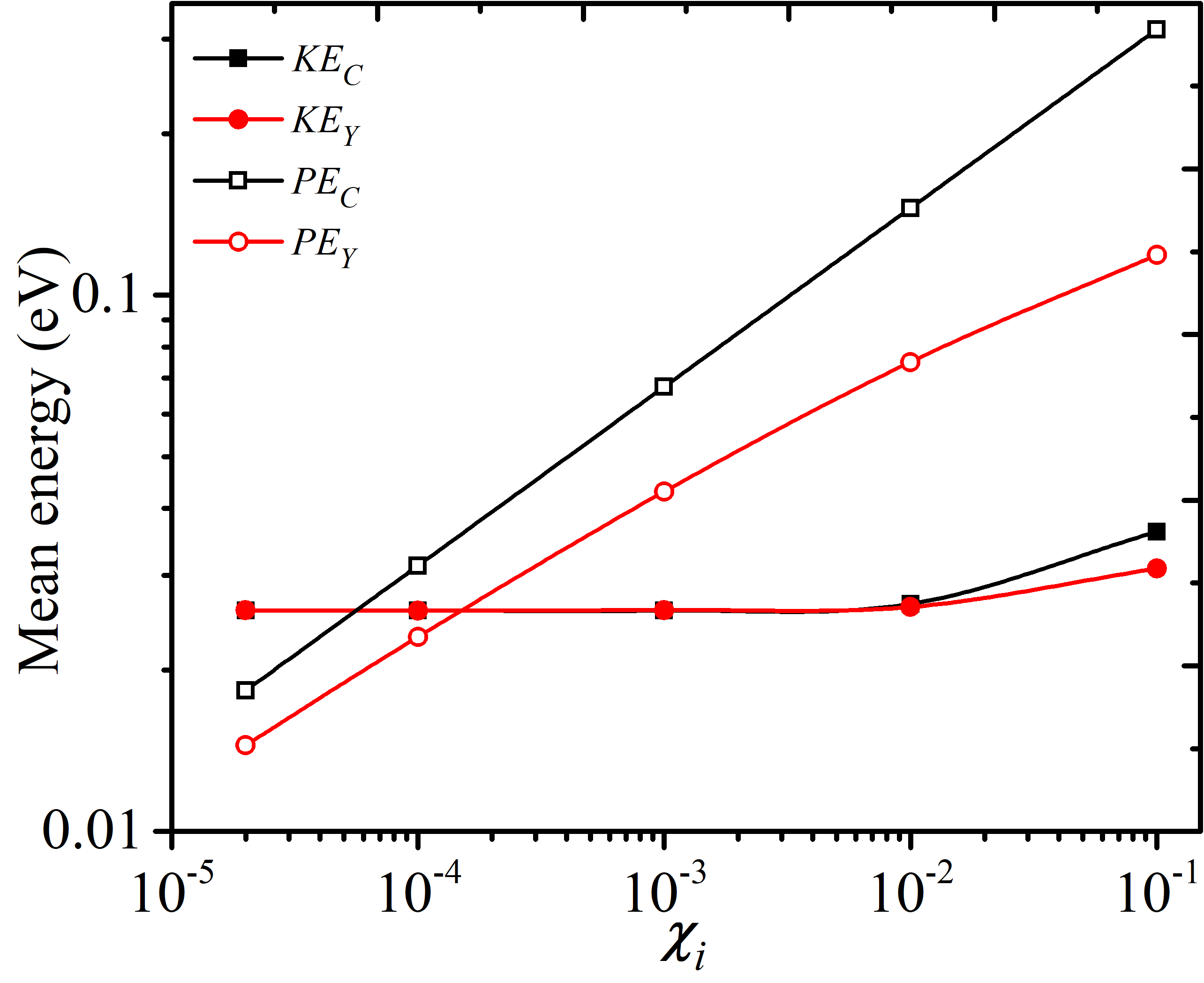}
         \caption{Mean potential ($PE$) and mean kinetic energies ($KE$) for ions for both potentials.}
\end{figure}

\begin{figure}
         \centering
         \includegraphics[width= 8 cm, height = 6.8 cm]{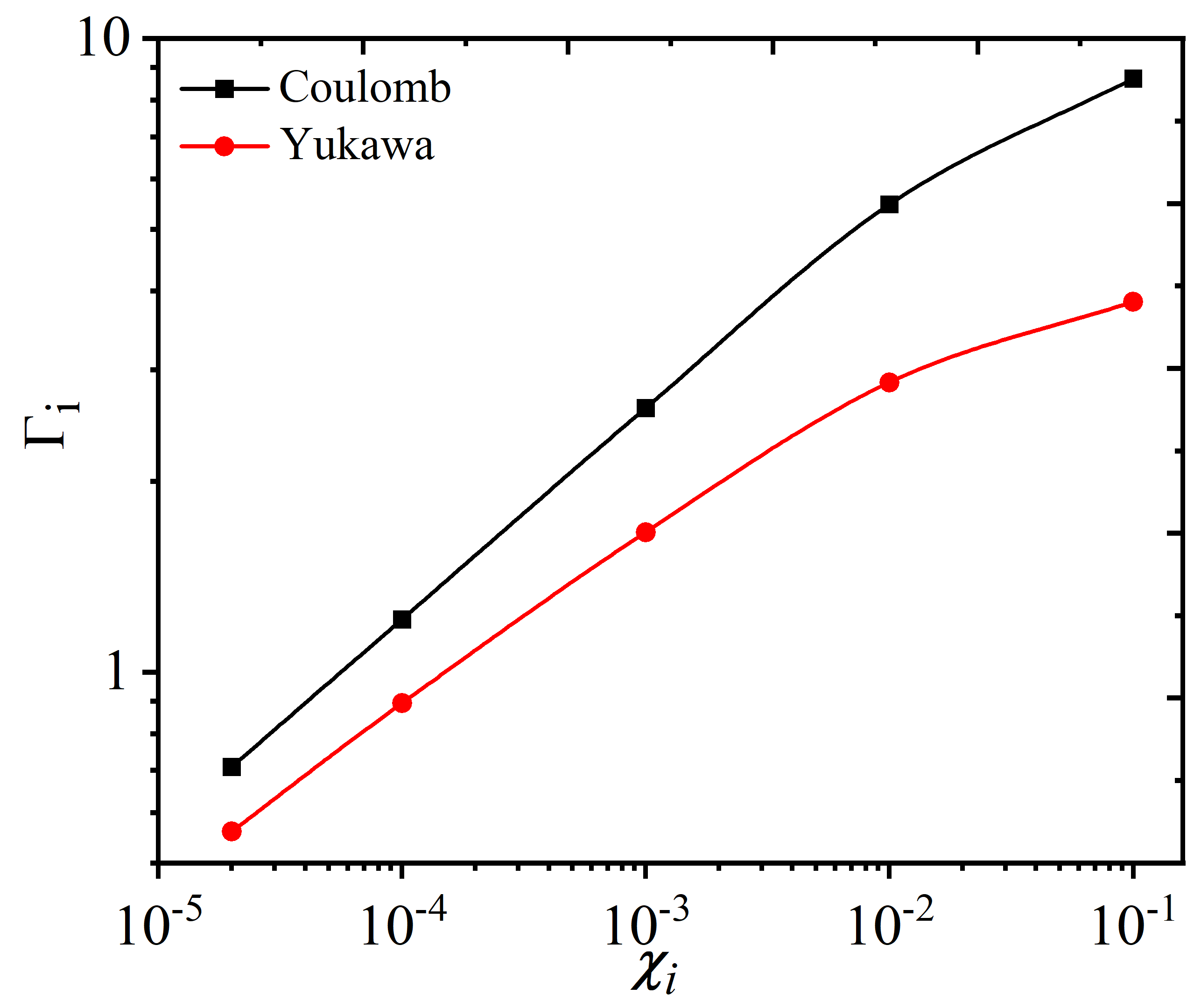}
         \caption{Ion coupling parameter ($\Gamma_{i}$) versus the ionization fraction ($\chi_{i}$) for Coulomb and Yukawa potentials. The solid lines are only to guide the eyes.}
\end{figure}
\begin{figure*}
\begin{subfigure}
  \centering
  \includegraphics[width= 8.15 cm, height = 6.9 cm]{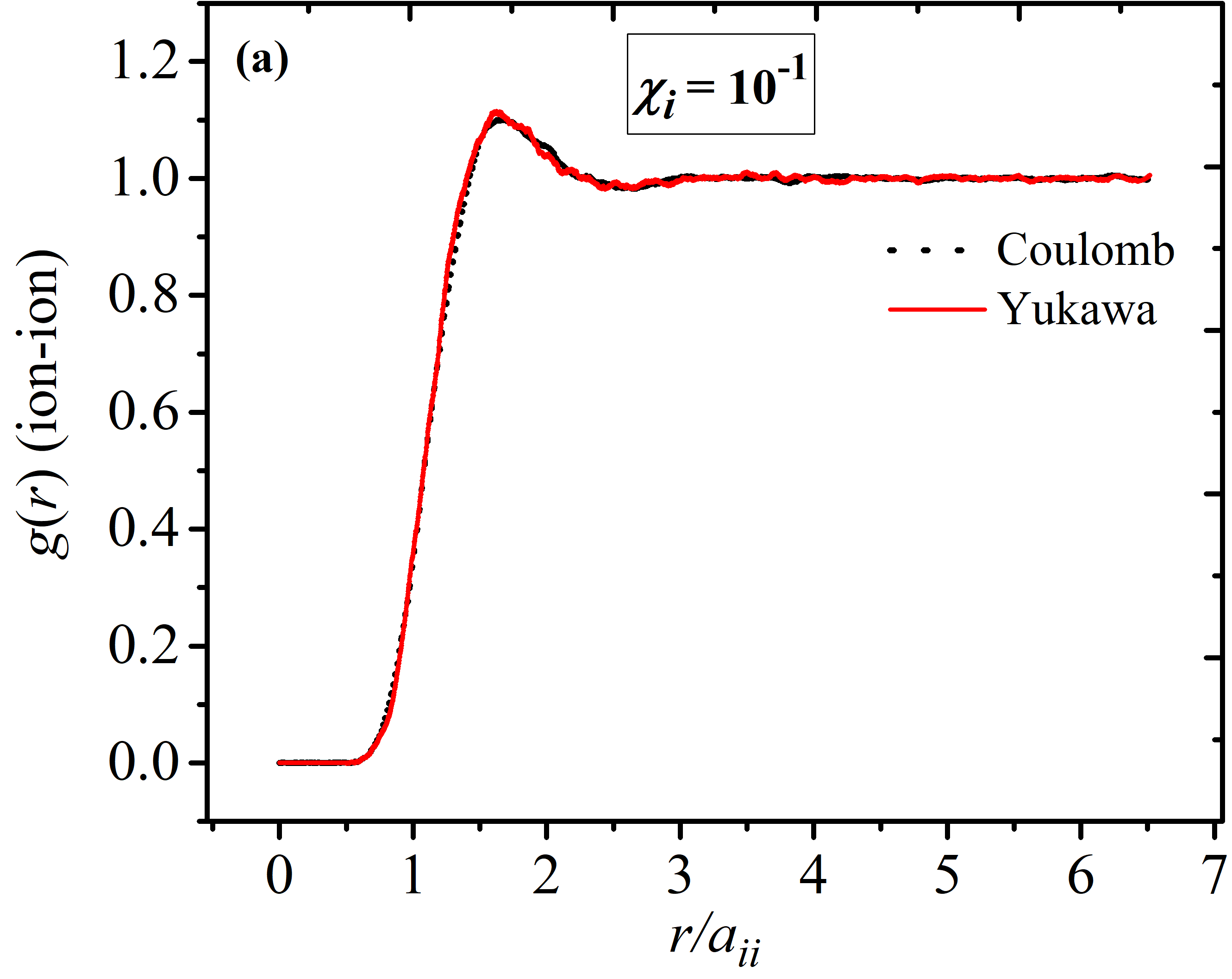}  
  \label{fig:sub-first}
\end{subfigure}
\begin{subfigure}
  \centering
  \includegraphics[width= 8.15 cm, height = 6.9 cm]{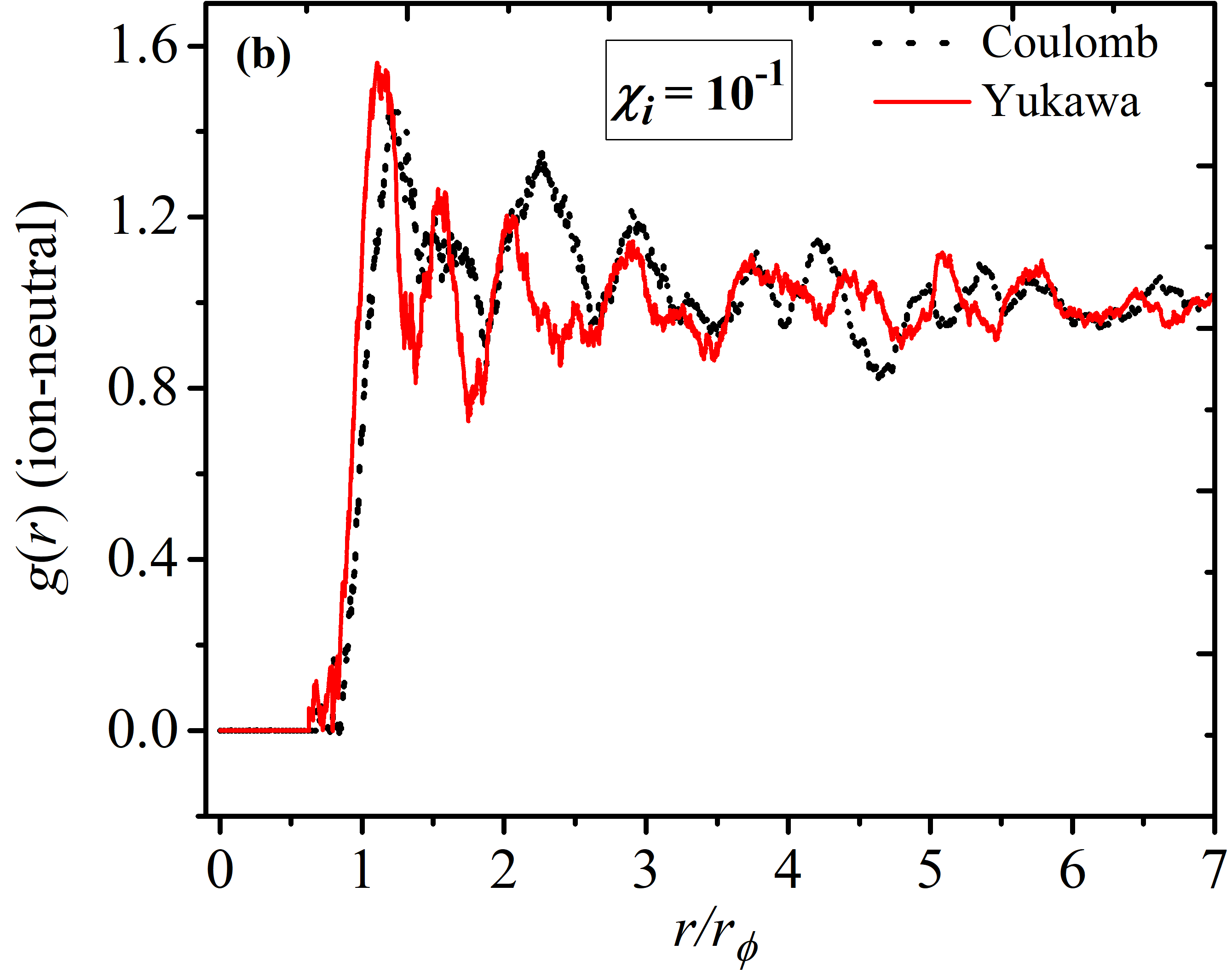}  
  \label{fig:sub-second}
\end{subfigure}
\caption{Radial distribution function for (a) ion-ion pair and (b) ion-neutral pair, for ionization fraction ($\chi_i) = 10^{-1}$.   }
\label{fig:fig}
\end{figure*}

\begin{figure*}
\begin{subfigure}
  \centering
  \includegraphics[width= 8.15 cm, height = 6.9 cm]{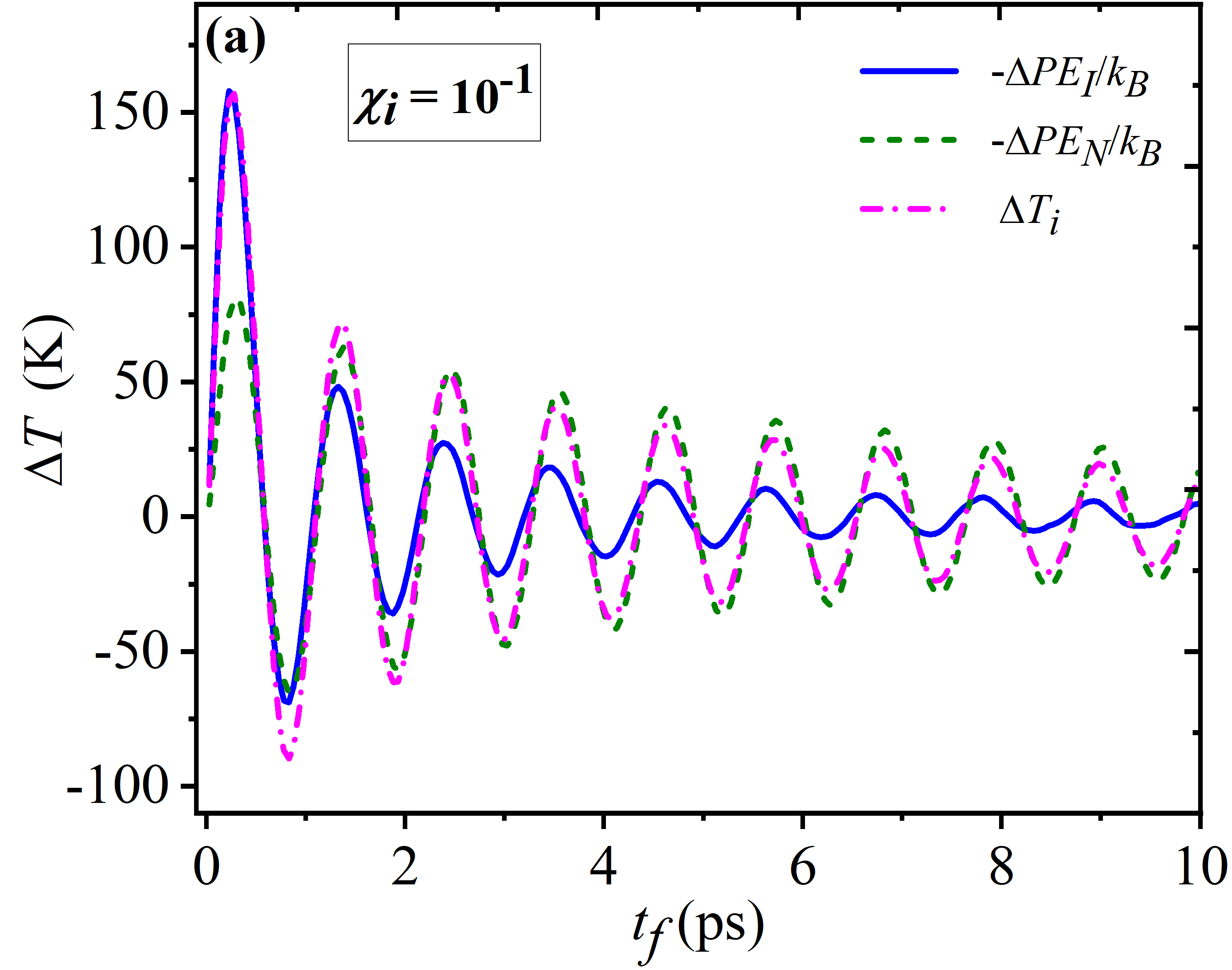}  
  \label{fig:sub-first}
\end{subfigure}
\begin{subfigure}
  \centering
  \includegraphics[width= 8.15 cm, height = 6.9 cm]{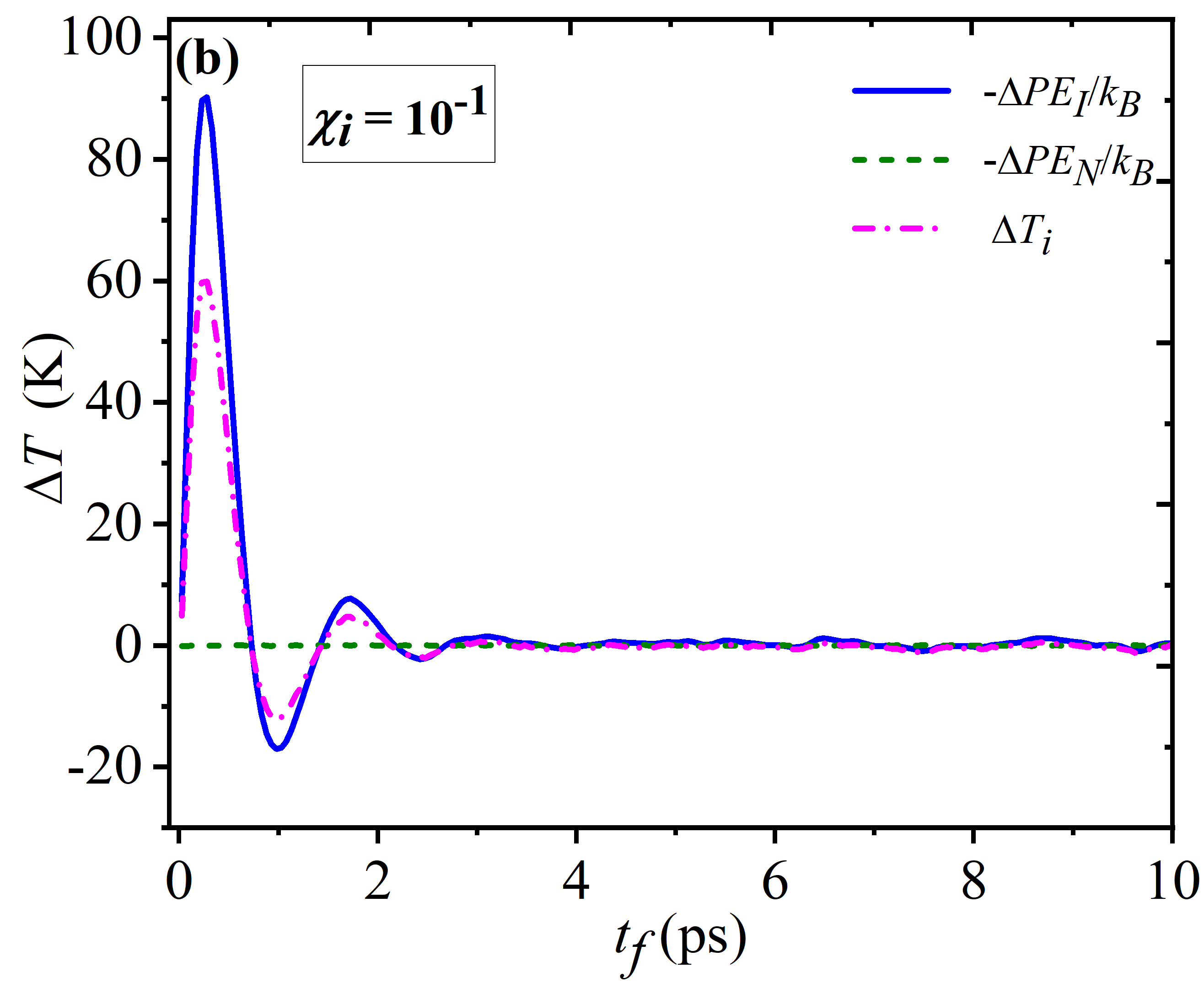}  
  \label{fig:sub-second}
\end{subfigure}
\caption{Change in temperature ($\Delta T$) versus time for (a) Coulomb potential and (b) Yukawa potential, for ionization fraction ($\chi_i) = 10^{-1}$. }
\label{fig:fig}
\end{figure*}


To study the ion dynamics during the DIH mechanism, the mean squared displacement for the ions (MSD = $\frac{1}{N_i}\sum_{j=1}^{N_i} [r_j(t) - r_j(0)]^2$, where $r_j(0)$ and $r_j(t)$ are the displacements of the $j^{th}$ ion at an initial time and a later time $t$, respectively) is calculated. MSD is calculated independently at different initial times in two scenarios: (1) during the non-equilibrium (NE) phase, as soon as the ions are introduced in the system, and (2) during the equilibrium (EQ) phase ($T_i = T_g$), for both potentials. Figure 3 shows the MSD data for $\chi_i = 10^{-1}$. It is observed that for the NE case, after a short ballistic regime (MSD $\propto t^a, a = 2$), the MSD shows a sub-diffusive regime with $a \sim 0.8$, before transitioning to the diffusive regime with $a = 1$ for both potentials. The time corresponding to the end of the ballistic regime also corresponds to $T_{i_\_max}$. Thereafter, the sub-diffusive regime starts and lasts till $T_i$ approaches $T_g$. Such a sub-diffusive regime is missing for the EQ scenario and for $\chi_i \le 10^{-4}$, even in the NE scenario. This observation confirms the fact that the sub-diffusive behavior is exclusive to the DIH mechanism and it indicates the non-thermal transport in the system. A similar observation has been reported for ultracold neutral plasma (UNP), where the ion velocity distribution during the DIH is non-thermal in nature due to the influence of nearest neighbors \cite{bergeson2011density}. This phenomenon is also observed in biological specimens, where the particle is crowded by its neighbors, which hinders its random motion \cite{luo2023simulation}.

\subsection{Ion coupling parameter}

The screening effect due to the electrons also has an effect on the system's energetics. The mean potential energy due to ion-ion pair interaction for Coulomb potential is given by,
\begin{equation}
PE_C = \frac{q^2}{4\pi\epsilon_{0}a_{ws}}, 
\end{equation}
where $a_{ws}$ is the Weigner-Seitz radius, which is calculated from the values of $n_i$ ($a_{ws}$ = $(3/4\pi n_i)^{1/3}$).
For Yukawa potential, the mean potential energy due to ion-ion pair interaction is given by,
\begin{equation}
PE_Y = \frac{q^2 e^{-\kappa a_{ws}}}{4\pi\epsilon_{0}a_{ws}}.
\end{equation} 
The mean kinetic energy ($KE$) for both potentials is $k_BT_i$. $T_i$ depends on the chosen potential at the equilibrium condition. Subscripts ``C" and ``Y" are used to refer to Coulomb and Yukawa potentials, respectively.
Figure 4 shows the variation of mean $PE$ and $KE$ with $\chi_i$ for both potentials. It is evident that the mean $PE$ for Yukawa potential is lower than that for Coulomb potential, which provides an energetically more favorable condition for the plasma system. Meanwhile, the mean $KE$ remains almost constant for both potentials except for the $\chi_i = 10^{-1}$ case.

The screening effect on the system's energetics is quantified through the coupling parameter of the ions ($\Gamma_{i}$).
For Coulomb potential, $\Gamma_{iC} = PE_C/KE_C$ (=$q^2/{4\pi\epsilon_{0}a_{ws}k_B T_i}$) and for Yukawa potential $\Gamma_{iY} = PE_Y/KE_Y$ (= ${q^2 e^{-\kappa a_{ws}}}/{4\pi\epsilon_{0}a_{ws}k_B T_i}$) \cite{vishnyakov2007coupling}.

It is observed that $\Gamma_{i}$ decreases with a decrease in $\chi_i$ due to the reduced number of ions for the lower values of $\chi_i$. The values of $\Gamma_{i}$ for Coulomb potential are higher than those for Yukawa potential due to the screening effect (cf. Figure 5). This finding agrees with the observations for the ultracold neutral plasmas \cite{lyon2013limit,ott2014coupling}.

\subsection{Internal plasma structure}

To probe into the internal structure of the plasma, the radial distribution function ($g(r)$ = ${V n(r)}/{4N\pi{r^2}\Delta{r}})$, where $n(r)$ is the number of particles situated from a reference particle at a distance of $r$ and $r+\Delta r$) for various pairs, such as ion-ion and ion-neutral, is computed at equilibrium condition for both potentials. The curves are shown for $\chi_i = 10^{-1}$ in Figure 6. For the ion-ion pair (cf. Figure 6(a)), the $g(r)$ curves for both potentials overlap each other. The values of $g(r)$ are negligible for $r/a_{ii} \le 0.54$. These curves attain a peak ($\sim 1.1$) at $r/a_{ii}$ = 1.65 and then remain constant with the value of 1 throughout. The peak height decreases and almost becomes 1 for lower values of $\chi_i$ (not shown). It signifies that the ions do not form any structures in the system \cite{mishra2022temperature}, and the electron screening does not impact the internal structure of the ions. For the ion-neutral pair (cf. Figure 6(b)), $g(r)$ is negligible for $r < r_{\phi}$. $r_{\phi}$ is the radius at which the repulsive core in the ion-neutral interaction potential acts (cf. eqn. (4)). $g(r)$ for the ion-neutral pair shows multiple peaks for both potentials and attains the primary peak at $r/ r_{\phi}$ = 1.23 and 1.10 for Coulomb and Yukawa potentials, respectively. It is observed that the peak widths are smaller for the Yukawa potential. Furthermore, the multiple secondary peaks indicate the formation of possible shell structures among the ions and neutrals \cite{yadav2023structure}. The position and width of the peaks dictate the ion-neutral interaction energy. The primary peak for the Coulomb potential falls in the region where the potential energy for the ion-neutral interactions is negative with a higher magnitude. A more detailed discussion of the potential energy of an ion due to the ion-neutral interaction can be found in Section IV and Appendix A. For lower values of $\chi_i$, the primary peak height increases for both potentials, with distinctive secondary peaks (not shown). 


\section{Discussion}

To understand the potential energy contribution to the DIH and oscillations in the $T_i$ curve, a model is developed. According to the DIH hypothesis, the increase in ion kinetic energy is attributed to the decrease in their potential energy. Mathematically, for an individual ion, it can be written as
\begin{equation}
    \Delta (k_B T_i) = - (\Delta PE_{T}),
\end{equation}
where $PE_{T}$ is the total pairwise potential energy experienced by an ion due to other ions and neutral particles. Therefore, $PE_T = PE_I + PE_N$, where $PE_I$ is the potential energy due to other ions which can be either due to Coulomb potential ($PE_C$) or Yukawa potential ($PE_Y$) and $PE_N$ is the ion potential energy due to the neutrals.

For two arbitrary initial ($t$) and final ($t+\Delta t  = t_f$) time instances, eqn. 8 can be written as,

\begin{multline}
    \Delta T = \underbrace{T_{i}(t+\Delta t) - T_{i}(t)}_{\Delta T_{i}} \\
    = \frac{1}{k_B} ((PE_{I} (t) - PE_{I} (t+\Delta t)) + \\
    (PE_{N} (t) - PE_{N} (t+\Delta t))),
\end{multline}
where $\Delta T$ is the change in temperature.
For Coulomb potential, the expression for $\Delta T$ becomes,
\begin{multline}
    = \underbrace{\frac{q^2}{4\pi\epsilon_{0}k_B} [\frac{1}{r(t)}-\frac{1}{r(t+\Delta t)}]}_{-\Delta PE_I/k_B} +  \\
   \underbrace{ \frac{q^2\alpha_R a_B^3}{8\pi\epsilon_{0}}[(\frac{r_{\phi}^8}{r^{12}(t)} - \frac{1}{r^4(t)})-(\frac{r_{\phi}^8}{r^{12}(t+\Delta t)} - \frac{1}{r^4(t+\Delta t)})]}_{-\Delta PE_N/k_B}.
\end{multline}

Similarly, for Yukawa potential, the expression for $\Delta T$ becomes,
\begin{multline}
    = \underbrace{\frac{q^2}{4\pi\epsilon_{0}k_B} [\frac{e^{\kappa r(t)}}{r(t)}-\frac{e^{\kappa r(t+\Delta t)}}{r(t+\Delta t)}]}_{-\Delta PE_I/k_B} +  \\
   \underbrace{ \frac{q^2\alpha_R a_B^3}{8\pi\epsilon_{0}}[(\frac{r_{\phi}^8}{r^{12}(t)} - \frac{1}{r^4(t)})-(\frac{r_{\phi}^8}{r^{12}(t+\Delta t)} - \frac{1}{r^4(t+\Delta t)})]}_{-\Delta PE_N/k_B}.
\end{multline}
 $\Delta T_{i}$ (cf. eqn. (9)) is the change in ion temperature in the time interval $\Delta t$ and is acquired from the simulations directly, where $\Delta t$ is chosen as 0.05 ps. $\Delta T$ due to the change in potential energies (R.H.S of eqns. (10) and (11)) is acquired from the simulation by taking the collective potential energies of all the ions and then averaging their values by dividing with $N_i$. 
$\Delta T$ is plotted against $t_f$ in Figure 7 for both potentials for $\chi_i = 10^{-1}$, showing  the contribution of different pairwise interactions to $\Delta T$. It is observed that for Coulomb potential, the initial rise of $T_i$ due to DIH is attributed to the $PE_I$ contribution. As the ions start to cool down, the contribution of $PE_I$ diminishes, and $PE_{N}$ increases. The rapid decrease in $PE_I$ with time is caused by the increased distance between the ions to attain a minimum potential energy state among themselves. For Yukawa potential, $\Delta T_i$ is primarily due to the $PE_I$ contribution and the $PE_N$ contribution is negligible throughout. Furthermore, it is found that, for Coulomb potential, the values of $PE_{N}$ are negative and large, whereas, for Yukawa potential, these are negative but negligibly small. Upon analyzing the ion-neutral interaction potential, it is understood that the ion-neutral interaction energy is negative and significant only for a small range of separation distance for an ion-neutral pair ($r/r_{\phi} \sim 1 - 2.4$) (cf. Figure A.1). For $r/r_{\phi} > 2.4$, the values of the interaction potential are negative but negligible. This shows that apart from the separation distance of the ion-neutral pair, the number of neutrals and their arrangement around an ion also play a crucial role in determining the values of $PE_{N}$.   

The damped oscillations in the $\Delta T$ curve at long timescales ($>$ 2 ps) are only observed for Coulomb potential. The frequency and the amplitude of these oscillations decrease with a decrease in $\chi_i$ due to lower ion densities (not shown in the graph). Furthermore, for the Yukawa potential, as $\kappa$ decreases and approaches the value 0, the oscillations appear and are comparable to those shown for the Coulomb potential (not shown graphically).

\section{Conclusion}
Atmospheric pressure He plasma at room temperature is modeled for $\chi_i = 10^{-1}$ - $10^{-5}$ employing molecular dynamics simulations, where two potentials are used to describe the ion-ion interaction; Coulomb and Yukawa. Yukawa potential incorporates the screening effect caused by the background electrons.

It is found that the DIH, which is prominent for $\chi_i \ge 10^{-3}$, is more significant for the Coulomb potential. The electron screening in the Yukawa potential efficiently limits the ion and gas heating in the system. The maximum ion temperatures for the weakly ionized plasmas ($\chi_i \le 10^{-4}$) are almost similar for both potentials. During the non-equilibrium phase ($T_i \ne T_g$) caused by the DIH, the $T_i$ curve for the Coulomb potential shows an oscillatory behavior for $\chi_i \ge 10^{-3}$. The oscillations are very much damped for the Yukawa potential. For weakly ionized plasmas ($\chi_i \le 10^{-4}$), such oscillations are absent for both potentials. During this non-equilibrium phase, the ions show a sub-diffusive behavior for $\chi_i \ge 10^{-3}$, which is more prominent for the Coulomb potential. The sub-diffusive behavior is exclusively associated with the DIH mechanism. 

The decrease in the potential energy due to the ion-ion interaction contributes to the increase in $T_i$ during the DIH process for both Coulomb and Yukawa potential cases. The contribution to the change in ion temperature by the ion-neutral interactions is dominant for Coulomb potential.

In general, the potential energy due to the ion-neutral interactions is attractive for both potentials and the magnitude is higher for Coulomb potential. Combined with the $g(r)$ data for the ion-neutral pair, it is understood that a higher number of neutrals are arranged near an ion for Coulomb potential, and the ion-neutrals form a shell-like structure for both potentials for all the values of $\chi_i$.

 To conclude, it is more scientific to include the screening effect in the simulations for the atmospheric pressure plasma systems, as this effect dictates the plasma dynamics, energetics, and internal structure more realistically. The mean potential energy in a Yukawa system is lower, which is energetically favourable for the stability of the system. Although the Coulomb potential is easier to implement with long-range effects and is computationally less expensive, the Yukawa potential may be important for ascertaining more physical results for the given experimental conditions. However, the Yukawa potential needs to be investigated further to ascertain how it guides the ion-neutral structural arrangements at room-temperature plasmas and to extend the current finding to high and low gas temperatures at atmospheric pressure, which may be taken up in the future works.

 \section{ACKNOWLEDGEMENT}
 The authors (S.S. Mishra and S. Bhattacharjee) sincerely thank the HPC facility at IIT Kanpur and the funding agency DST-SERB India for the research grant CRG/2022/000112.

\appendix
\section{Appendix}
Figure A.1 shows the values of the interaction potential ($\phi_{ind}$) and force ($F_{ind}$) for the ion-neutral interactions for $\chi_i = 10 ^{-1}$. For $r/r_{\phi} \le 1.14$, the values for $F_{ind}$ are positive (repulsive). $\phi_{ind}$ attains minimum at $r/r_{\phi} = 1.14$, and the values of $F_{ind}$ becomes negative (attractive). 
\setcounter{figure}{0}
\renewcommand{\figurename}{FIG.A.}

\begin{figure} 
         \centering
         \includegraphics[width= 8.7 cm, height = 6.5 cm]{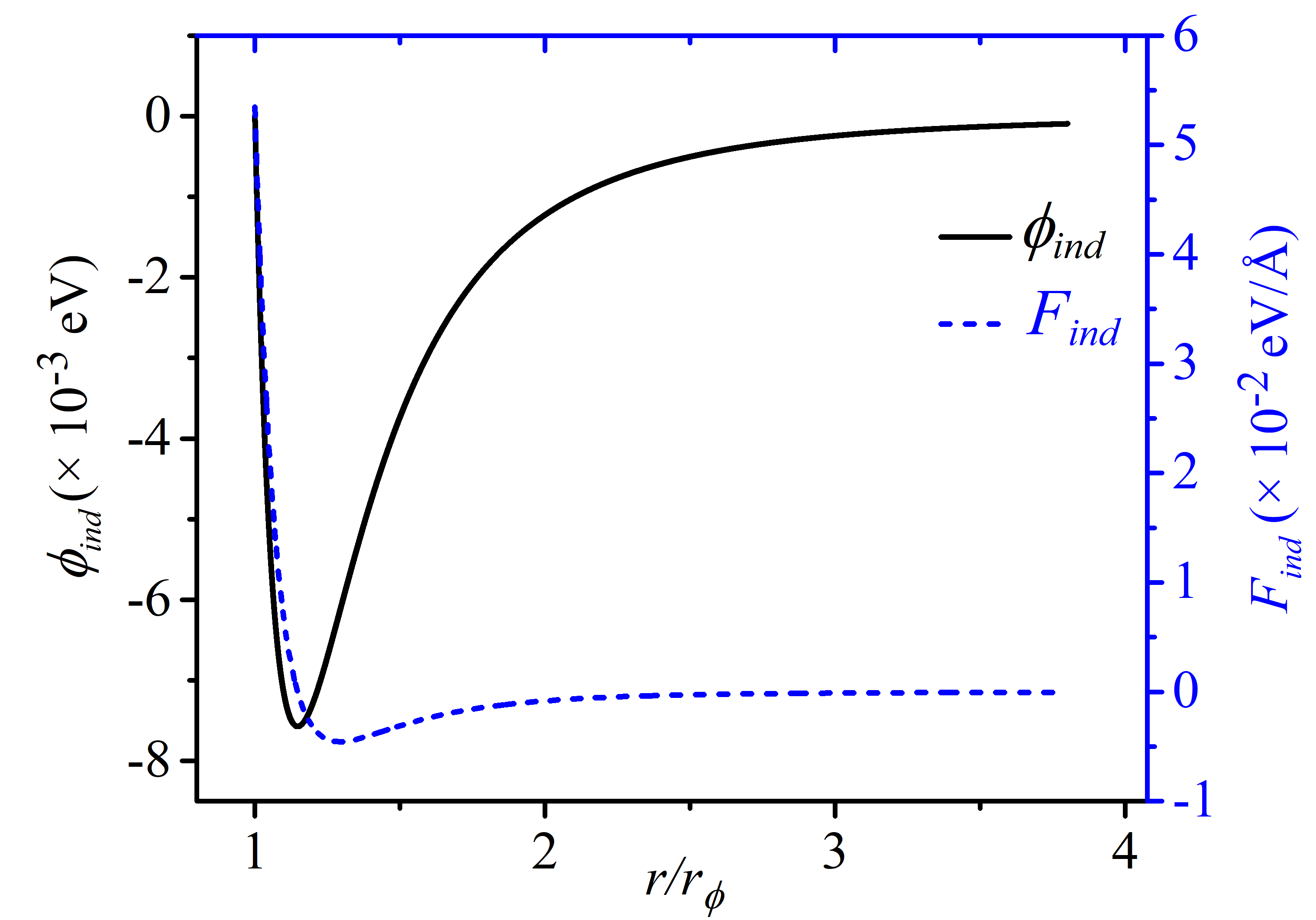}
         \caption{Ion neutral interaction potential and force values for ionization fraction, $\chi_i = 10^{-1}$.} 
         \label{fig:figA1}
         
\end{figure}

\bibliography{apssamp}

\end{document}